# THE COMPETITION BETWEEN STAGGERED FIELD AND ANTIFERROMAGNETIC INTERACTIONS IN $CuGeO_3$:Fe

S.V Demishev [1)], A.V.Semeno [1)], A.A.Pronin [1)], N.E.Sluchanko [1)], N.A.Samarin [1)], H.Ohta [2)], S.Okubo [2)], M.Kimata [3)], K.Koyama [4)], M. Motokawa [4)]

[1)] *Low Temperatures and Cryogenic Engineering Department, General Physics Institute of RAS, Vavilov street, 38, 119991 Moscow, Russia*

[2)] *Molecular Photoscience Research Center, Kobe University, 1-1 Rokkodai, Nada, Kobe 657-8501, Japan*

[3)] *The Graduate School of Science and Technology, Kobe University, 1-1 Rokkodai, Nada, Kobe 657-8501, Japan*

[4)] *Institute for Materials Research, Tohoku University, Sendai 980-8577, Japan*

## Abstract

The EPR spectra along different crystallographic axes for single crystals of $CuGeO_3$ containing 1% of Fe impurity have been studied in the frequency range 60-360 GHz at temperatures 0.5-30 K. The analysis based on the Oshikawa-Affleck (OA) theory suggests that the temperature dependences of the line width and *g*-factor are formed as a result of the competition between interchain antiferromagnetic interactions and staggered Zeeman energy. It is found that staggered magnetic moments in $CuGeO_3$:Fe are located predominantly along ***b*** axis.

**Key words:** $CuGeO_3$, EPR, Oshikawa-Affleck theory

## 1. Introduction.

Up to very nearest past the general way of understanding of electron paramagnetic resonance in 1D quantum antiferromagnetic (AF) spin chains was based on the exchange narrowing (EN) theory [1-3]. However the application of the results [1-2] initially obtained and well justified for the 3D single spin EPR problem to 1D system of the strongly coupled spins is not straightforward. The EPR in 1D AF spin chain is essentially a collective mode and hence the real spin dynamics may be different from the simple random Gaussian process [4-5]. As a consequence the EPR line shape will change the substantially in the range $T < J/k_B$, where correlation between spins in chain cannot be neglected (hereafter $J$ denotes the absolute value of exchange integral).

In the present paper we will consider the case of doped $CuGeO_3$. In the last decade this material have attracted much attention as the only one example of inorganic spin-Peierls compound and thus representing a good experimental realization of quasi-1D quantum spin system. In the terms of EN theory the EPR characteristics of $CuGeO_3$ have been examined in [6]. The obtained results suggest that EN theory allows to describe experimental data only in the range $T > 130$ K. The latter value correlate well with the $J/k_B$=120 K characteristic to $CuGeO_3$ [7]. Moreover, down to ~15 K the EPR line have retained the same lorentzian shape [6].

An alternative approach to EPR in 1D S=1/2 AF spin chains have been suggested recently by Oshikawa and Affleck [8]. The Oshikawa-Affleck (OA) theory predicts lorentzian form of the EPR line, which width is controlled by two contributions, namely staggered field (SF) and exchange anisotropy (EA). The justification of the applicability of the OA theory to the case of doped $CuGeO_3$ has been provided in [9]. It was shown that $CuGeO_3$ doped with 1% of Fe might serve as a good experimental example where both effects of EA and SF can be studied [9]. The possible mechanisms responsible for the appearance of SF in $CuGeO_3$ doped with magnetic impurities were considered in [10]. However the detailed checking of the OA theory in a wide range of frequencies, magnetic fields and temperatures and for the various alignments of the external magnetic field **B** with respect to crystallographic axes is still missing for $CuGeO_3$:Fe and the present work is aimed on the experimental solution of this problem.

## 2. Theoretical background.

In the presence of the exchange anisotropy and staggered field the line width and *g*-factor in the OA theory are given by [8-9]

$$W(T)=W_{EA}(T) + W_{SF}(T)=A\cdot T + C\cdot h^2\cdot T^{-2}, \quad (1a)$$

$$g(T)=g_0+g_{SF}(T)= g_0 + D\cdot h^2\cdot T^{-3}, \quad (1b)$$

where $h$ denotes magnitude of the staggered field, and $A$, $C$ and $D$ stands for the quantities having weak temperature dependence and calculated in [8]. We wish to emphasize that the low temperature growth of $W(T)$ and $g(T)$ is caused by staggered field and these parameters are strongly correlated, i.e. the values $C$ and $D$ are differ by numerical coefficient [8]. Following the ansatz suggested in [9] we will use Eqs. (2) to define the OA function

$$f_{OA}(T)=W(T)/[\Delta g(T)\cdot T]\equiv A^{*}\cdot(T^{3}/h^{2}) + C_{OA}, \quad (2)$$

where $\Delta g(T)= g(T)-g_0$ , $A^{*}=A/D$ and $C_{OA}$ is the universal constant in the OA theory, which does not depend on exchange integral and SF magnitude: $C_{OA}=1.99\cdot k_B/\mu_B$ [8-9].

Equation (2) suggests that in 1D case $f_{OA}\rightarrow C_{OA}$ when $T\rightarrow 0$. This behavior changes dramatically when a possibility of 3D AF ordering is taken into account [11], as long as approaching of the Neel point $T_N$ from above leads to the line width divergence $W\sim(T-T_N)^{-3/4}$ [12]. A possible correction to the *g*-factor may be attributed to the difference between local and external magnetic field, i.e. $\Delta g(T)\sim\chi(T)$ (here $\chi$ denotes magnetic susceptibility). Thus in 3D AF model the OA function acquires the form [11]

$$f_{OA}(T)\sim[T\cdot\chi(T)\cdot(T-T_N)^{4/3}]^{-1}, \qquad (3)$$

and $f_{OA}(T\rightarrow T_N)\rightarrow\infty$. Consequently the analysis of the experimental data by computing of the OA function allows discriminating EPR line broadening caused by AF correlations and staggered field.

**3. Results and discussion.**

The EPR spectra for single crystals of $CuGeO_3$ containing 1% of Fe impurity have been studied in the frequency range 60-360 GHz for the case when external magnetic field ***B*** was parallel to ***a*** crystallographic axis. Both cavity (60-110 GHz) and quasi-optical (100-360 GHz) techniques have been utilized. The measurements for $\boldsymbol{B}||\boldsymbol{b}$ and $\boldsymbol{B}||\boldsymbol{c}$ have been performed in 60 GHz cavity spectrometer. Experiments were carried out mainly in the temperature range 1.8-30 K; however for $\boldsymbol{B}||\boldsymbol{a}$ and 60 GHz cavity we were able to extend temperature range down to 0.5 K using $He^3$ cryostat. The details about samples preparation and quality control may be found elsewhere [13].

The experimental EPR spectra for $CuGeO_3$:Fe (fig. 1) consist of a single line corresponding to resonance on chains of $Cu^{2+}$ ions in agreement with the published data [9,13].The line shape was found to be lorentzian within the experimental accuracy. The quality of the spectra has allowed to find a full set of spectroscopic data (line widths $W$, $g$-factors, integrated intensities $I$ ) for the whole temperature-frequency domain studied. The $I(T)$ data are presented in [14] and suggest that doping with 1% of Fe impurity (S=2) of $CuGeO_3$ induces in high quality single crystals a strong disorder in magnetic subsystem of $Cu^{2+}$ (S=1/2) quantum spin chains and leads to a complete damping of both spin-Peierls and Neel transitions down to 0.5 K. Therefore the OA theory may be applicable in the frequency and temperature diapasons studied and below we will consider $W(T)$ and $g(T)$ behavior.

The temperature dependences of the line width demonstrate low temperature minimum (see fig. 1 and fig. 2). This is in agreement with the OA theory predictions for the case when both EA and SF effects take place (Eq. (1a)). It follows from the insets in fig. 1-2 that the low temperature increase of the *g*-factor occurs simultaneously (Eq. (1b)). Note that at 60 GHz for $CuGeO_3$:Fe the above picture is qualitatively the same for various orientations of magnetic field with respect to crystal axes (fig. 1) whereas in pure $CuGeO_3$ line width always decreases with lowering temperature [15] and the temperature dependence of the *g*-factor is different for $\boldsymbol{B}||\boldsymbol{a}$ $\boldsymbol{B}||\boldsymbol{b}$ and $\boldsymbol{B}||\boldsymbol{c}$ (see solid lines in inset at fig. 1). The data in fig. 2 suggests that increase of frequency (and hence the resonant field) induces enhancement of the line width, which may caused by the field dependent staggered field contribution (in equation (1) the magnitude of $h$ increases with $B$ [8-9]). However at the same time for frequency range $\nu \geq 100$ GHz the low temperature growth of the *g*-factor apparently slows down (inset in fig. 2).

In order to clarify mechanisms causing broadening of the line width and temperature correction to the *g*-factor we have computed OA functions from the data shown in fig. 1 and fig. 2. The result is presented in fig. 3 and fig. 4 respectively; solid lines are corresponding to fits using Eq. (2). In calculations the value $\Delta g(T)= g(T)- g(30\text{ K})$ have been used. It is visible that OA theory provides reasonable description of the experimental data at 60 GHz for various orientations of magnetic field in the interval $T > 2$ K. Interesting that for $\boldsymbol{B}||\boldsymbol{a}$ and $\boldsymbol{B}||\boldsymbol{c}$ the magnitude of the coefficient at $T^3$ in Eq. (2) is almost equal and noticeably bigger than in case $\boldsymbol{B}||\boldsymbol{b}$ (fig. 3). As long as in OA theory $A^*$ does not depend on magnetic field orientation, this result suggests that staggered component of the magnetic moment is located predominantly along $\boldsymbol{b}$ axis. In $CuGeO_3$ structure this direction corresponds to the strongest interchain interactions [7].

For $T < 2$ K the experimental OA function starts to increase when temperature is lowered and apparently deviates from the predictions of the OA theory. This behavior may probably reflect approaching AF transition temperature in qualitative agreement with Eq. (3). Hence the observed non-monotonous dependence of the $f_{OA}(T)$ may reflect transition from predominantly 1D behavior of $Cu^{2+}$ chains ($T$ >2 K) to beginning of the 3D AF ordering ($T$ <2 K). Further studies in the temperature range $T$ <0.5 K are required to check this hypothesis.

The above consideration suggests that AF interactions in $CuGeO_3$:Fe may become stronger for higher frequencies (magnetic field). The data in fig. 4 indicate that for ν≥100 GHz at low temperature the OA asymptotic (2) is no longer valid and the observed $f_{OA}(T)$ can be understood in the 3D AF ordering model (Eq. (3)).

In conclusion, we have shown that in $CuGeO_3$:Fe the observed temperature dependences of the line width and $g$-factor are formed as a result of the competition between interchain antiferromagnetic interactions in $CuGeO_3$ and staggered Zeeman energy. The importance of this phenomenon has been recently demonstrated theoretically in [16]. The AF interactions becomes more pronounced at $T$<2 K and in frequency range ν≥100 GHz. In the region of applicability of the OA theory the analysis of the experimental data suggests that staggered magnetic moments in $CuGeO_3$:Fe are located predominantly along ***b*** axis.

Authors acknowledge support from Russian Science Support Foundation, programmes “Physics of Nanostructures” and “Strongly Correlated Electrons” of RAS and grants RFBR 04-02-16574 and INTAS 03-51-3036. This work was partly supported by Hyogo Science and Technology Association and also by Grant-in-Aid for Science Research on Priority Areas (No. 13130204) from the Ministry of Education, Culture, Sports, Science and Technology of Japan.

**References**

1. J.H. Van Vleck, Phys. Rev., **74**, 1168 (1948)
2. P.W. Anderson, P.R. Weiss, Rev. Mod. Phys., **25**, 269 (1953)
3. R. Kubo, K. Tomita, J. Phys. Soc. Jpn., **9**, 888 (1954)

4. R.E.Dietz, F.R.Merritt, R. Dingle, D. Hone, B.G. Silbernagel, P.M. Richards, Phys.Rev.Lett., **26**, 1186 (1971)
5. G.F.Reiter, J.P. Boucher, Phys. Rev. B, **11**, 1823 (1975)
6. R.M. Eremina, M.V.Eremin, V.N.Glazkov, H.-A. Krug von Nidda, A.Loidl, Phys. Rev. B, **68**, 014417 (2003)
7. L.P.Regnault, M.Ain, B.Hennion, G.Dhalenne, A.Revcolevschi, Phys. Rev. B, **53**, 9, 5579 (1996)
8. M.Oshikawa, I.Affleck. Phys. Rev. Lett., **82**, 5136 (1999); cond-mat/0108424
9. S.V.Demishev, Y.Inagaki, H.Ohta, S.Okubo, Y.Oshima, A.A.Pronin, N.A.Samarin, A.V.Semeno, N.E.Sluchanko, Europhys. Lett., **63**, 446 (2003)
10. S.V.Demishev, cond-mat/0405366
11. S.V. Demishev, A.V. Semeno, N.E. Sluchanko, N.A.Samarin, A.A. Pronin, Y. Inagaki, S Okubo, H. Ohta, Y. Oshima, L.I. Leonyuk, Physics of the Solid State, 46, 2238 (2004); cond-mat/0404639
12. H.Mori, K.Kawasaki, Prog.Theor.Physics, **28**, 971 (1962)
13. S.V.Demishev, R.V.Bunting, L.I.Lleonyuk, E.D.Obraztsova, A.A.Pronin, N.E.Sluchanko, N.A.Samarin, S.V.Terekhov, JETP Letters, **73**, 31 (2001)
14. S.V.Demishev, A.V.Semeno, A.A.Pronin, N.E.Sluchanko, N.A.Samarin, H.Ohta, S.Okubo, M.Kimata, K.Koyama, M.Motokawa, A.V.Kuznetsov, cond-mat/0408242
15. S.V.Demishev, A.V.Semeno, N.E.Sluchankko, N.A.Samarin, A.N.Vasil'ev, L.I.Leonyuk, JETP, **85**, 943 (1997)
16. M.Sato, M.Oshikawa, Phys.Rev. B, **69**, 054406 (2004)

**Figure captions.**

Fig. 1. Temperature dependences of line width and $g$-factor for various orientations of magnetic field measured at frequency 60 GHz.

Fig. 2. Temperature dependences of line width and $g$-factor for $\mathbf{B} || \mathbf{a}$ at various frequencies.

Fig. 3. The OA function calculated from the data in fig. 1. Points- experiment, solid lines- best fits using Eq. (2). Dashed line- universal constant in the OA theory. Dash-dot line is a guide to the eye.

Fig. 4. . The OA function calculated from the data in fig. 2. The signs are the same as in fig. 3.

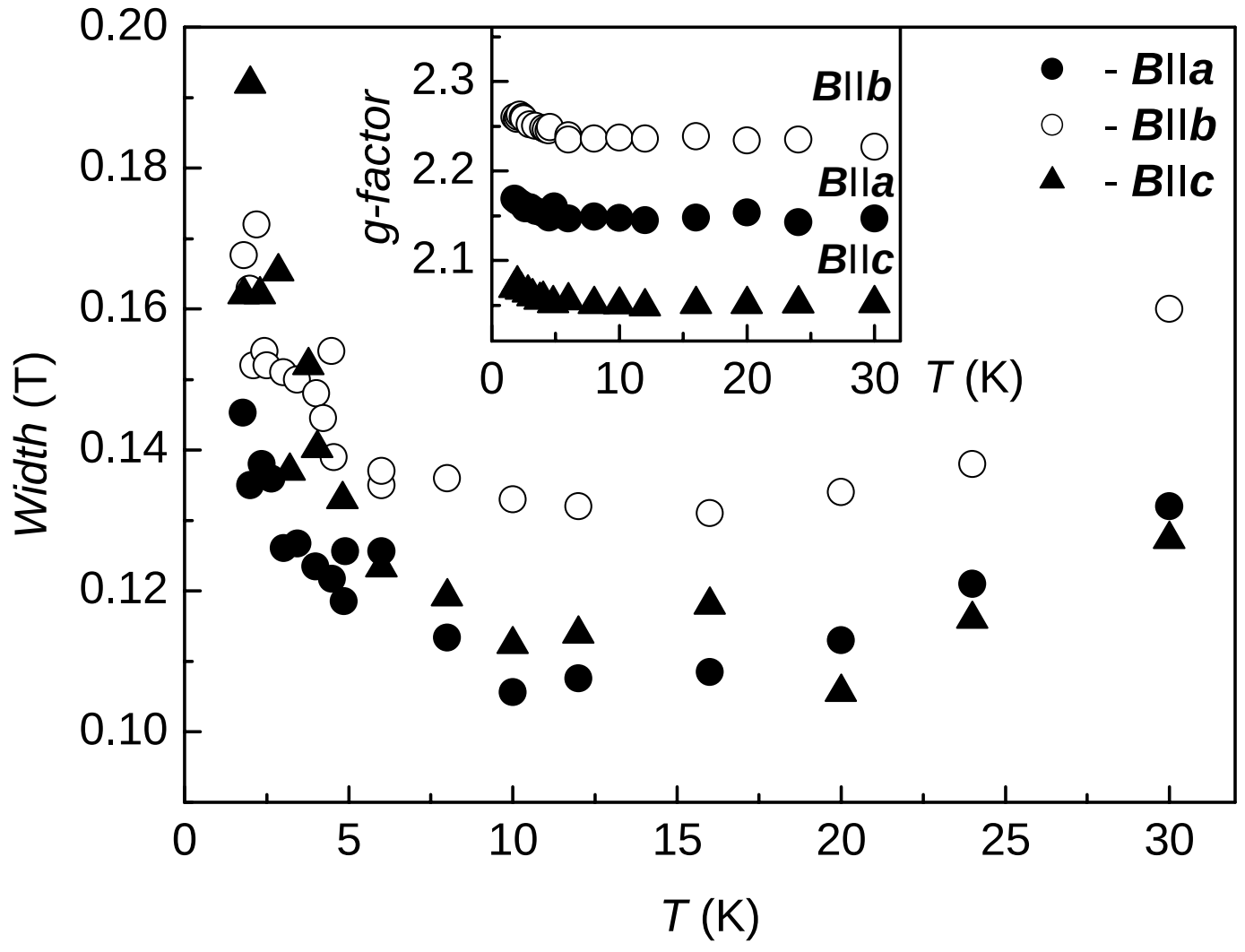

Fig.1 S.V.Demishev et al.

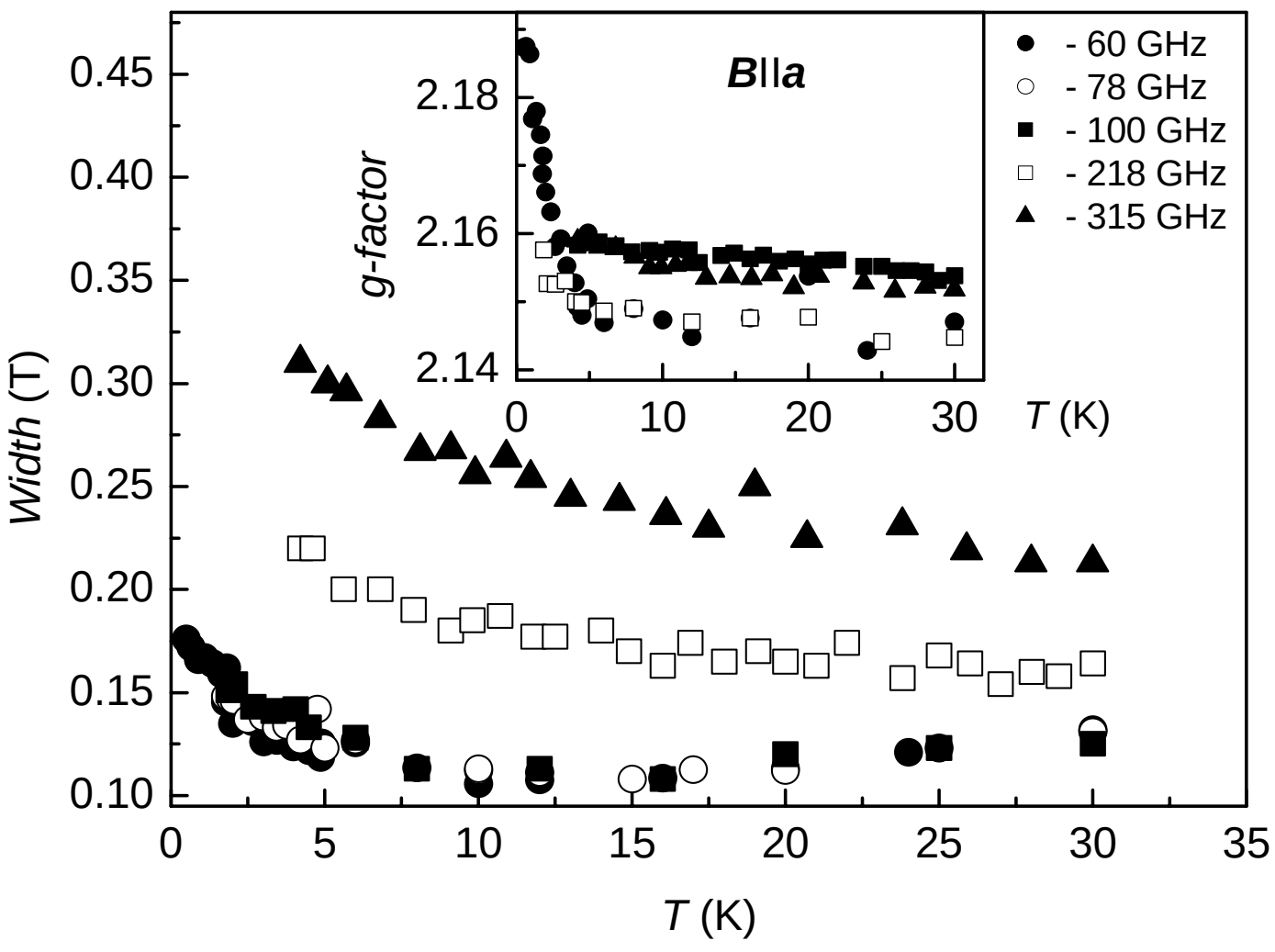

Fig.2 S.V.Demishev et al.

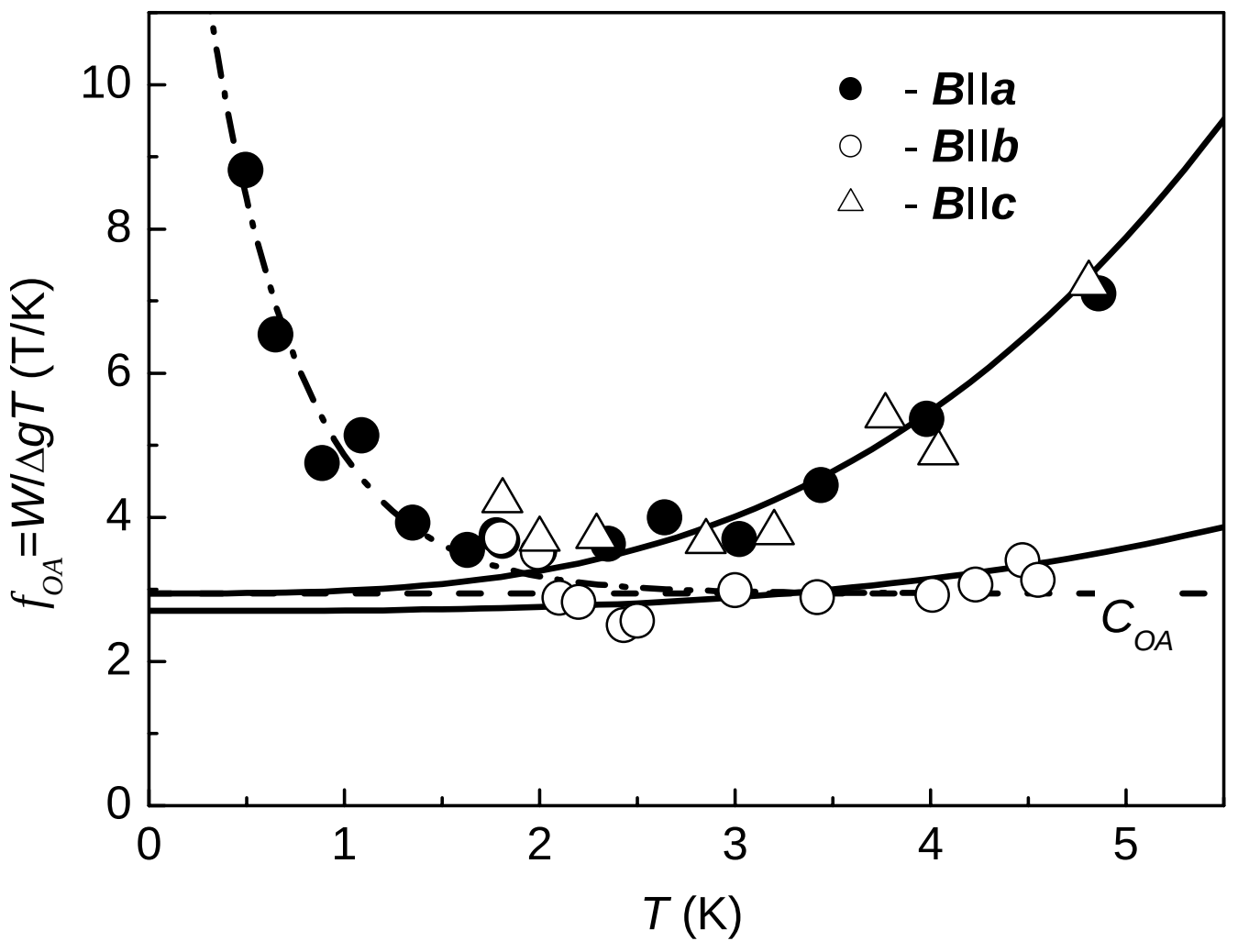

Fig.3 S.V.Demishev et al.

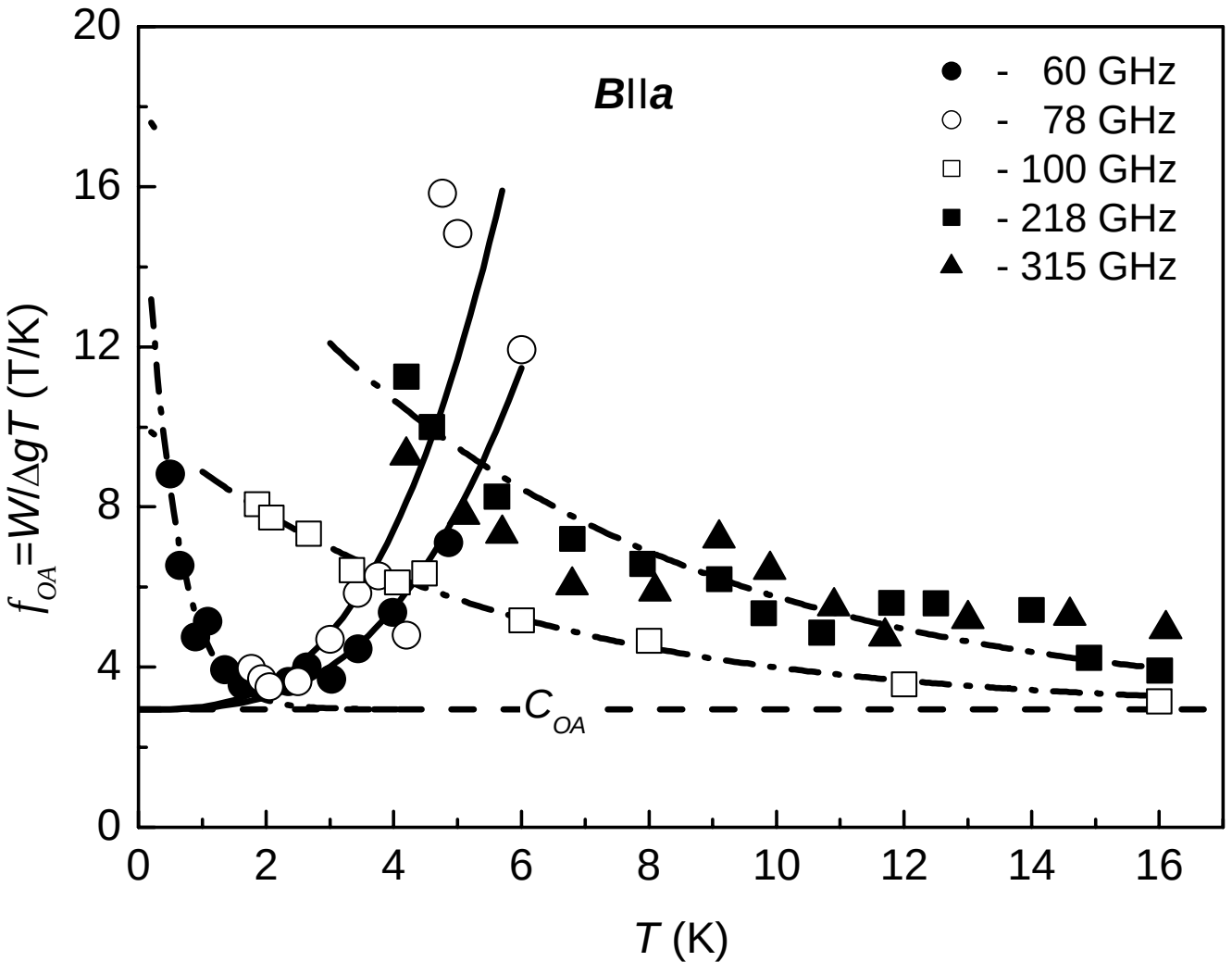

Fig.4 S.V.Demishev et al.